\documentclass[11pt, a4paper]{article}
\pdfoutput=1
\usepackage{jcappub}

\usepackage{booktabs}
\usepackage{multirow}
\usepackage{amssymb}
\usepackage[export]{adjustbox}
\usepackage{floatrow}
\newfloatcommand{capbtabbox}{table}[][3.5in]
\newfloatcommand{capbfigbox}{figure}[][2.3in]
\usepackage{blindtext}
\usepackage{amsmath}
\usepackage{booktabs}
\usepackage{aas_macros}

\newcommand{\ie}{i.e.~}

\def\lsim{\mathrel{\raise.3ex\hbox{$<$\kern-.75em\lower1ex\hbox{$\sim$}}}}
\def\gsim{\mathrel{\raise.3ex\hbox{$>$\kern-.75em\lower1ex\hbox{$\sim$}}}}

\begin{document}

\hspace*{110mm}{\large \tt FERMILAB-PUB-18-069-A}

\vskip 0.2in

\title{Millisecond Pulsars, TeV Halos, and Implications For The Galactic Center Gamma-Ray Excess}

\author{Dan Hooper$^{a,b,c}$}\note{ORCID: http://orcid.org/0000-0001-8837-4127}
\emailAdd{dhooper@fnal.gov}
\author{and Tim Linden$^d$}\note{ORCID: http://orcid.org/0000-0001-9888-0971}
\emailAdd{linden.70@osu.edu}

\affiliation[a]{Fermi National Accelerator Laboratory, Center for Particle Astrophysics, Batavia, IL 60510}
\affiliation[b]{University of Chicago, Department of Astronomy and Astrophysics, Chicago, IL 60637}
\affiliation[c]{University of Chicago, Kavli Institute for Cosmological Physics, Chicago, IL 60637}
\affiliation[e]{Ohio State University, Center for Cosmology and AstroParticle Physics (CCAPP), Columbus, OH  43210}

\abstract{Observations by HAWC indicate that many young pulsars (including Geminga and Monogem) are surrounded by spatially extended, multi-TeV emitting regions.  It is not currently known, however, whether TeV emission is also produced by recycled, millisecond pulsars (MSPs). In this study, we perform a stacked analysis of 24 MSPs within HAWC's field-of-view, finding between 2.6--3.2$\sigma$ evidence that these sources are, in fact, surrounded by TeV halos. The efficiency with which these MSPs produce TeV halos is similar to that exhibited by young pulsars. This result suggests that several dozen MSPs will ultimately be detectable by HAWC, including many ``invisible'' pulsars without radio beams oriented in our direction. The TeV halos of unresolved MSPs could also dominate the TeV-scale diffuse emission observed at high galactic latitudes. We also discuss the possibility that TeV and radio observations could be used to constrain the population of MSPs that is present in the inner Milky Way, thereby providing us with a new way to test the hypothesis that MSPs are responsible for the Galactic Center GeV excess.}

\maketitle

\section{Introduction}
\label{sec:introduction}

The High Altitude Water Cherenkov (HAWC) Observatory has recently presented their first source catalog~\cite{Abeysekara:2017hyn}, as well as a second analysis focusing on the nearby pulsars Geminga and Monogem~\cite{newhawc}. Unexpectedly, these sources exhibit an extended profile of multi-TeV emission, corresponding to a physical extent of at least $\sim$25~pc. The extended nature of this emission strongly favors an inverse Compton origin and demonstrates that these pulsars inject approximately 10\% of their total spindown power into very high-energy electron-positron pairs~\cite{Hooper:2017gtd}. 

Further inspection of the 2HWC Source Catalog~\cite{Abeysekara:2017hyn}, as well as HESS observations of pulsar wind nebulae~\cite{Abdalla:2017vci}, suggests that ``TeV halos'' are very likely not limited to Geminga and Monogem, but are instead a feature common to most pulsars. More specifically, of the 39 sources contained in the 2HWC Catalog, 16 are associated or potentially associated with a known pulsar (compared to an expected 2.7 chance associations)~\cite{Linden:2017vvb}. Furthermore, among the young (100-400 kyr) pulsars in the Australia Telescope National Facility (ATNF) Catalog~\cite{2005AJ....129.1993M} within HAWC's field-of-view and with the largest predicted gamma-ray flux (assuming a Geminga-like TeV efficiency), 6 out of 11 have potential associations with HAWC sources~\cite{Linden:2017vvb}. One of these six sources (HAWC J0543+233) was even predicted~\cite{Linden:2017vvb} to produce an observable TeV halo prior to its reported observation.\footnote{The detection of this TeV halo was reported in the following astronomer's telegram: \url{www.astronomerstelegram.org/?read=10941}} All indications are that HAWC's source catalog is dominated by pulsars, that TeV halos are present around most, if not all, young pulsars, and that sub-threshold TeV halos produce the majority of diffuse TeV emission~\cite{Linden:2017vvb,Linden:2017blp}.

In this paper, we examine whether TeV halos also exist around recycled pulsars with millisecond-scale periods. Although the 2HWC Catalog does not contain any millisecond pulsar (MSP) candidates,\footnote{Although the HAWC Collaboration points out that the MSP J1950+2414 is located only $\sim$$0.3^{\circ}$ from the 2HWC source J1949+244, the spindown power and distance of this source make this association highly unlikely.} this does not necessarily indicate that this class of objects are not surrounded by TeV halos. In particular, only a handful of MSPs within HAWC's field-of-view have a spindown flux (which we define throughout this study as the spindown power divided by the distance squared) high enough to produce a TeV halo that would currently be detectable by HAWC. Furthermore, it is generally expected that MSPs do indeed generate such emission~\cite{Venter:2015gga,Bednarek:2016gpp,Venter:2015oza}, as the modeling of their light curves favor the abundant production of multi-TeV electron-positron pairs~\cite{Venter:2015gga}. On the other hand, the high-surface brightness of TeV halos indicates that cosmic-ray diffusion is inhibited in the vicinity of these sources, and it is unknown whether MSPs are capable of generating such conditions.

The question of whether MSPs are surrounded by TeV halos has important implications for gamma-ray astrophysics, and such halos could provide a new way to identify and constrain the emission from MSPs. Additionally, it has been suggested that a large population of unresolved MSPs may be responsible for the GeV gamma-ray excess observed from the region surrounding the Galactic Center. If TeV halos in fact exist around this pulsar population, the associated TeV and radio synchrotron emission will likely be observable, providing a new way to confirm or constrain this hypothesis.

The remainder of this article is structured as follows. In Sec.~\ref{model}, we briefly review what is currently known about the physics of TeV halos. In Sec.~\ref{analysis}, we describe a simple analysis of HAWC data that finds between 2.6 and 3.2$\sigma$ evidence in favor of TeV halos around MSPs, with a similar efficiency to that of Geminga and Monogem. In Sec.~\ref{implications}, we discuss some of the astrophysical implications of TeV halos around MSPs. In Sec.~\ref{IG}, we consider the TeV halos of MSPs within the context of observations of the Inner Galaxy. In particular, we find that if MSPs are confirmed to be surrounded by TeV halos, then a pulsar population that is capable of generating the GeV excess would also be expected to saturate or exceed the TeV and/or radio synchrotron emission that is observed from this part of the sky. In Sec.~\ref{summary}, we summarize our results and conclusions.

\section{A Qualitative Model of TeV Halos}
\label{model}

While many of the details of the mechanism that produces bright, spatially extended TeV emission from pulsars are unknown, observations indicate that several components are necessary. First, the high luminosity of TeV halos indicates that a significant fraction (roughly 10\%) of the pulsar spindown power is converted into very high-energy electrons and positrons. In the case of young pulsars, this acceleration could occur either in the pulsar magnetosphere, or later as electrons are accelerated across the significant voltage drop in the termination shock of the pulsar wind nebula (PWN)~\cite{Gaensler:2006ua}. While dim PWN have been observed around some MSPs~\cite{Cheng:2006sm,2013AA...550A..39B}, MSPs with detected PWN have primarily been correlated with systems undergoing accretion from a binary companion. The detection of TeV halos among a significant fraction of MSPs would thus provide evidence for TeV electron/positron acceleration and escape from the pulsar magnetosphere. 

Second, the high surface brightness of TeV halos indicates that these very high-energy electron-positron pairs are confined to within $\sim$20~pc of the central pulsar~\cite{Abeysekara:2017hyn,newhawc}. In an environment such as the standard interstellar medium (characterized by a diffusion coefficient inferred from measurements of boron-to-carbon and other cosmic-ray secondary-to-primary ratios, $D\approx 4 \times 10^{28} \, (E_e/{\rm GeV})^{0.33}$ cm$^2/$s~\cite{GALPROPSite,Hooper:2017tkg}), a 10 TeV electron will diffuse a few hundred parsecs before losing most of its energy through synchrotron and inverse Compton processes. The angular distribution of the emission measured by HAWC indicates that the diffusion coefficient is two to three orders of magnitude smaller than the standard value in the regions surrounding Geminga and Monogem. This suggests that pulsars temporarily produce regions in which particle diffusion is significantly inhibited. In the case of young pulsars, these peculiar regions may potentially be correlated with either the associated supernova or with the star-formation regions in which many pulsars are born. If TeV halos with similar extension are observed surrounding MSPs, this would instead indicate that the pulsars themselves must provide the mechanism that inhibits local diffusion. 

It is worth stressing that neither of these effects is naively predicted by theory (and the second is unexpected), thus the observation of TeV halos surrounding MSPs would have the potential to significantly impact our understanding of both cosmic-ray acceleration and diffusion in and around these sources.

\section{Evidence for TeV Halos around Millisecond Pulsars}
\label{analysis}

Although the HAWC dataset is not fully accessible to those outside of the HAWC Collaboration, a public tool (see \url{data.hawc-observatory.org}) was recently made available that enables one to search the HAWC sky for evidence of point-like or extended sources. More specifically, this tool determines the test statistic (TS) of the hypothesis that there is a source in a given direction for either a point source with a spectral index of 2.7, or for an extended source with a spectral index of 2.0 and radial extension of either $0.5^{\circ}$, $1^{\circ}$ or $2^{\circ}$.  Although the results of such a tool will necessarily leave many interesting questions unanswered, we will employ it here in an effort to test whether MSPs are surrounded by TeV halos.

To this end, we consider a sample of MSPs within HAWC's field-of-view (declination between $-20^{\circ}$ and 50$^{\circ}$) selected by the value of their spindown flux (defined as spindown power divided by distance squared). We also require that each MSP in our sample is at least $2^{\circ}$ from every point-like source in the 2HWC catalog, and at least $2^{\circ}$ away from the edge of any extended source (using the source extensions reported in the 2HWC~\cite{Abeysekara:2017hyn}). Applying this criteria, we identify 24 MSPs in the ATNF catalog~\cite{2005AJ....129.1993M} with $\dot{E}/d^2 > 5\times 10^{33}$ erg/kpc$^2$/s (for comparison, Geminga has a value of $\dot{E}/d^2  \sim  5\times 10^{35} \, $erg/kpc$^2$/s). When available, we utilize distance measurements based on parallax, as opposed to those based on radio dispersion~\cite{2005AJ....129.1993M,1994ApJ...425L..41B,2005ApJ...634.1242N,2016AA...587A.109G,2012ApJ...756L..25D,2012ApJ...755...39V,2009MNRAS.400..951V,2011AIPC.1357...40H,2000ApJ...540L..41W,1994ApJ...426L..85A,TheFermi-LAT:2013ssa,2014ApJ...782L..38D}.

\begin{table*}
\label{Table1}
\begin{tabular}{|c|c|c|c|c|c|c|}
\hline 
PSR Name  & $\dot{E}$   & Distance  & $\dot{E}/D^2$ &Method & (TS)$^{1/2}$ & (TS)$^{1/2}$ \tabularnewline
  &  (erg/s) & (kpc)  &erg/kpc$^2$/s &  &(point-like) & (Geminga-like) \tabularnewline
\hline 
\hline 
J1400-1431  & $9.7\times 10^{33}$  & $0.28$ &$1.2\times 10^{35}$ & DM~\cite{2005AJ....129.1993M} & -1.17 & -1.65 \tabularnewline
\hline
J0034-0534 & $3.0\times 10^{34}$  & $0.54 \pm 0.10$ &$1.0^{+0.5}_{-0.3}\times 10^{35}$& DM~\cite{1994ApJ...425L..41B} & 1.40 & 0.40 \tabularnewline
\hline
J1737-0811  & $4.3\times 10^{33}$  & $0.21$ &$9.8\times 10^{34}$& DM~\cite{2005AJ....129.1993M} & 1.57 & -0.985 \tabularnewline
\hline
J1231-1411  & $1.8\times 10^{34}$  & $0.44\pm0.05$ & $9.3^{+2.5}_{-1.8}\times 10^{34}$&DM~\cite{2005ApJ...634.1242N} & -0.52 & -1.33 \tabularnewline
\hline 
J2214+3000  & $1.9\times 10^{34}$  & $0.59^{+0.66}_{-0.21}$&$5.5^{+7.7}_{-4.2} \times 10^{34}$ & P~\cite{2016AA...587A.109G} & -0.31 & 0.09 \tabularnewline
\hline
J1023+0038  & $9.8\times 10^{34}$  & $1.37^{+0.04}_{-0.03}$ &$5.2^{+0.3}_{-0.3} \times 10^{34}$& P~\cite{2012ApJ...756L..25D} & {\bf 2.18} & 1.63 \tabularnewline
\hline
J0030+0451 & $3.5\times 10^{33}$  & $0.28^{+0.10}_{-0.06}$ & $4.5^{+2.8}_{-2.0}\times 10^{34}$ & P~\cite{2012ApJ...755...39V} & -0.48 & {\bf 2.07} \tabularnewline 
\hline
J1843-1113  & $6.0\times 10^{34}$  & $1.26$ &$3.8\times 10^{34}$& DM~\cite{2005AJ....129.1993M} & 0.16 & 0.49 \tabularnewline
\hline
J1643-1224  & $7.4\times 10^{33}$  & $0.45^{+0.11}_{-0.07}$ &$3.7^{+1.5}_{-1.3}\times 10^{34}$& P~\cite{2009MNRAS.400..951V} & -0.45 & 0.73 \tabularnewline 
\hline
J0023+0923 & $1.6\times 10^{34}$  & $0.69^{+0.21}_{-0.11}$ &$3.4^{+1.4}_{-1.4}\times 10^{34}$& DM~\cite{2011AIPC.1357...40H} & 0.83 & 0.07 \tabularnewline
\hline
J1300+1240 & $1.9\times 10^{34}$  & $0.77^{+0.34}_{-0.18}$ &$3.2^{+2.3}_{-1.7}\times 10^{34}$& P~\cite{2000ApJ...540L..41W} & -0.26 & 0.68 \tabularnewline
\hline
J1744-1134  & $5.2\times 10^{33}$  & $0.42^{+0.03}_{-0.02}$ & $3.0^{+0.3}_{-0.2}\times 10^{34}$  & P~\cite{2009MNRAS.400..951V} & 0.13 & -0.79 \tabularnewline
\hline
J1959+2048  & $1.6\times 10^{35}$  & $2.49^{+0.16}_{-0.49}$ &$2.6^{+1.4}_{-3.0}\times 10^{34}$& DM~\cite{1994ApJ...426L..85A} & {\bf 2.54} & {\bf 2.54} \tabularnewline
\hline
J0337+1715 & $3.4\times 10^{34}$  & $1.30$ &$2.0\times 10^{34}$& DM~\cite{2005AJ....129.1993M} & 1.11 & 1.82 \tabularnewline
\hline
J1741+1351 & $2.3\times 10^{34}$  & $1.08^{+0.04}_{-0.05}$ &$2.0^{+0.2}_{-0.1}\times 10^{34}$& P~\cite{TheFermi-LAT:2013ssa} & -0.47 & -0.13 \tabularnewline 
\hline
J2017+0603  & $1.4\times 10^{34}$  & $0.83^{+0.60}_{-0.24}$ & $1.9^{+1.8}_{-1.3}\times 10^{34}$ & DM~\cite{2016AA...587A.109G} & -0.02 & 0.92 \tabularnewline
\hline
J2339-0533  & $2.3\times 10^{34}$  & $1.10$ &$1.9\times 10^{34}$& DM~\cite{2005AJ....129.1993M} & -1.35 & -2.07 \tabularnewline
\hline
J1939+2134  & $1.1\times 10^{36}$  & $7.7^{+7.7}_{-2.6}$ & $1.9^{+2.3}_{-1.4}\times 10^{34}$& P~\cite{2009MNRAS.400..951V} & {\bf 2.61} & {\bf 2.61} \tabularnewline
\hline
J0613-0200 & $1.3\times 10^{34}$  & $0.90^{+0.40}_{-0.20}$ &$1.6^{+1.0}_{-0.8}\times 10^{34}$& P~\cite{2012ApJ...755...39V} & 1.02 & {\bf 2.34} \tabularnewline
\hline
J1719-1438  & $1.6\times 10^{33}$  & $0.34$ &$1.4\times 10^{34}$& DM~\cite{2005AJ....129.1993M} & 0.27 & -0.44 \tabularnewline 
\hline
J1911-1114  & $1.2\times 10^{34}$  & $1.07$ &$1.0\times 10^{34}$& DM~\cite{2005AJ....129.1993M} & -0.48 & -0.76 \tabularnewline
\hline
J1745-0952  & $5.0\times 10^{32}$  & $0.23$ &$9.5 \times 10^{33}$& DM~\cite{2005AJ....129.1993M} & -1.40 & -2.29 \tabularnewline
\hline
J0218+4232  & $2.4\times 10^{35}$  & $6.3^{+8.0}_{-1.7}$ & $6.1^{+8.9}_{-5.0}\times 10^{33}$ & P~\cite{2014ApJ...782L..38D} & 0.36 & 0.36 \tabularnewline
\hline
J0557+1550  & $1.7\times 10^{34}$  & $1.83$ &$5.1\times 10^{33}$& DM~\cite{2005AJ....129.1993M} & -0.20 & -0.34 \tabularnewline
\hline 
\hline
\end{tabular}
\caption{The 24 millisecond pulsars in the HAWC field-of-view with $\dot{E}/d^2 > 5 \times 10^{33}$ erg/kpc$^2$/s and which are not located near another HAWC source. For each pulsar, we list the measured distance and error bars when available. We adopt parallax (P) measurements when possible, as those based on the dispersion measure (DM) rely on modeling of the traversed medium. In the two rightmost columns we list the test statistic (TS) found using the HAWC 2HWC Survey online tool for the case of a point-like gamma-ray source and a gamma-ray source with Geminga-like extension, respectively. The observation that 4 out of 24 of these sources exhibit (TS)$^{1/2} > 2.07$ (for the case of Geminga-like extension) has a chance probability of only 0.13\% (assuming the null hypothesis), corresponding to 3.2$\sigma$ evidence in favor of TeV halos around this collection of MSPs.}
\label{table1}
\end{table*}

In Table~\ref{table1}, we list the 24 MSPs in our sample and the values of their spindown power, distance, and spindown flux. We also list in this table the value of the test statistic (TS) calculated by the HAWC online tool for two source hypotheses. First, we simply test for the presence of a point-like source at the location of the MSP. Second, we test for the presence of an extended source with a physical extent most similar to that exhibited by Geminga and Monogem. More specifically, we use the point-like template for sources with $d > 2$ kpc, the $0.5^{\circ}$ template for $d=0.75-2$ kpc, the $1^{\circ}$ template for $d=0.375-0.75$ kpc, and the $2^{\circ}$ template for $d<0.375$ kpc. 

\begin{figure}
\includegraphics[width=4.60in,angle=0]{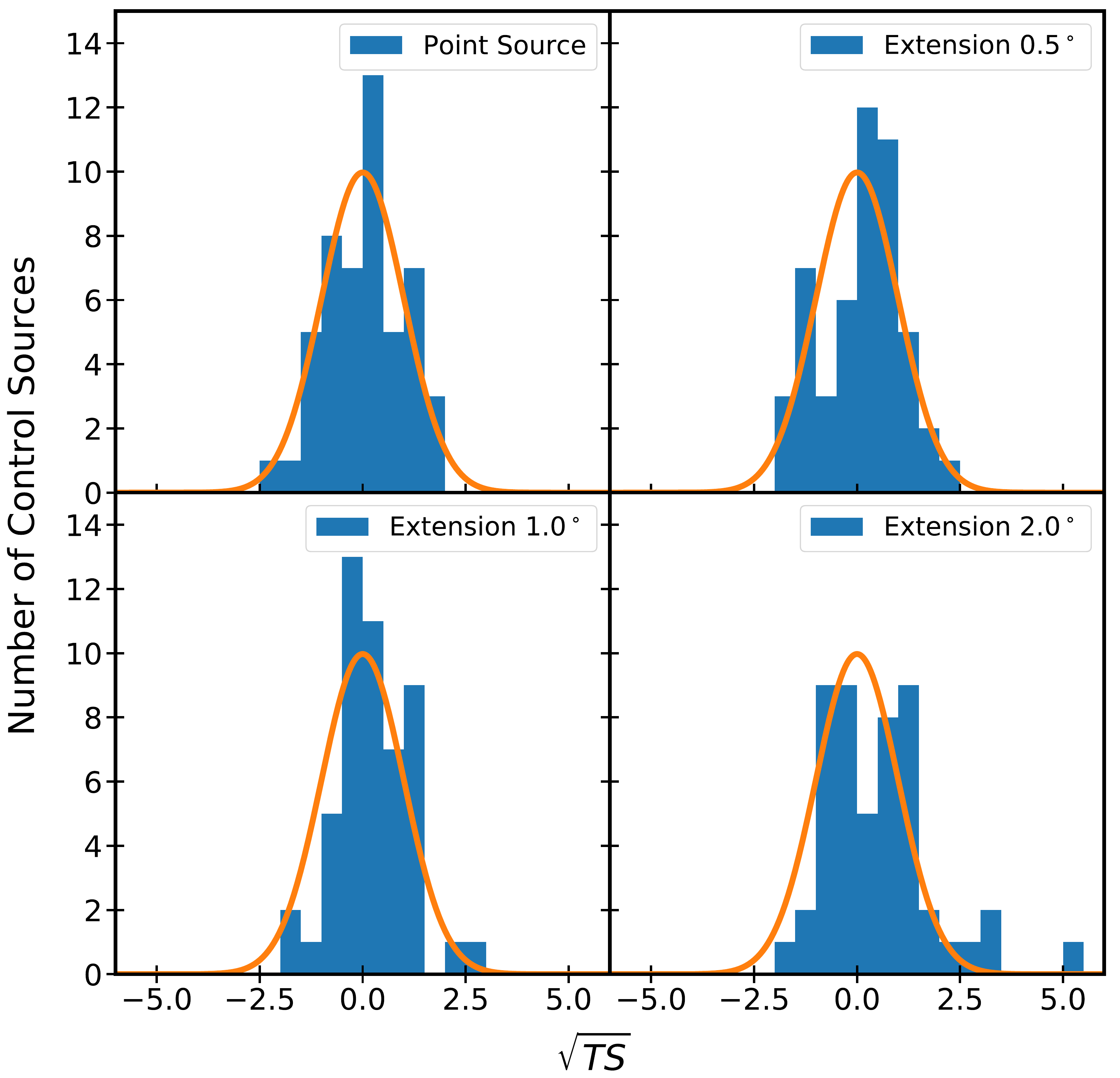}
\caption{The distribution of the test statistic (TS) found for the 50 MSPs in our control group.}
\label{control}
\end{figure}

Among these 24 sources, we find four which exhibit (TS)$^{1/2} \ge 2.07$ in favor of TeV halos with Geminga-like extension (and three with (TS)$^{1/2} \ge 2.18$ in favor of point-like TeV halos). If the TS distribution is described by a (two-sided) normal distribution, we calculate that this should occur with a chance probability of only 0.13\% (assuming the null hypothesis), corresponding to 3.2$\sigma$ evidence of TeV halos around this collection of MSPs. To further test whether the TS values are in fact normally distributed across the HAWC sky, we adopt a control group of 50 MSPs with no nearby HAWC sources and with the {\it lowest} values of $\dot{E}/d^2$ as reported in the ATNF catalog~\cite{2005AJ....129.1993M}, and show the resulting TS distributions in Fig.~\ref{control}. For the case of the point source template, the assumption that the TS values are described by a normal distribution is well-supported by the control group. In particular, among this collection of targets, the largest TS$^{1/2}$ obtained is 1.95, and 33 of these 50 sources yielded $|{\rm TS}| < 1$, entirely consistent with a normal distribution that includes no unexpectedly large tails. For the extended templates, the TS distribution of the control group sources appears somewhat broader, and skews slightly towards positive TS values. In order to take this information into account, we note that our model (based on the measured distances to the MSPs in our sample) contains three, nine, seven and five sources which utilize the point-like, $0.5^{\circ}$, $1^{\circ}$ and $2^{\circ}$ templates, respectively. Drawing from this collection of control group distributions, we find that only 0.46\% of the random selections produce four or more sources with a TS exceeding 2.07, corresponding to a significance of 2.6$\sigma$.\footnote{Our results are not highly sensitive to the precise choice of our TS cut. For example, the significance of four sources with TS$^{1/2} >2.0$ is 3.0$\sigma$ (2.2$\sigma$), and of six sources with TS$^{1/2} >1.5$ is 2.7$\sigma$ (2.6$\sigma$), adopting the point-like template (combined point-like and extended templates) for our control group.}

\begin{figure}
\includegraphics[width=4.60in,angle=0]{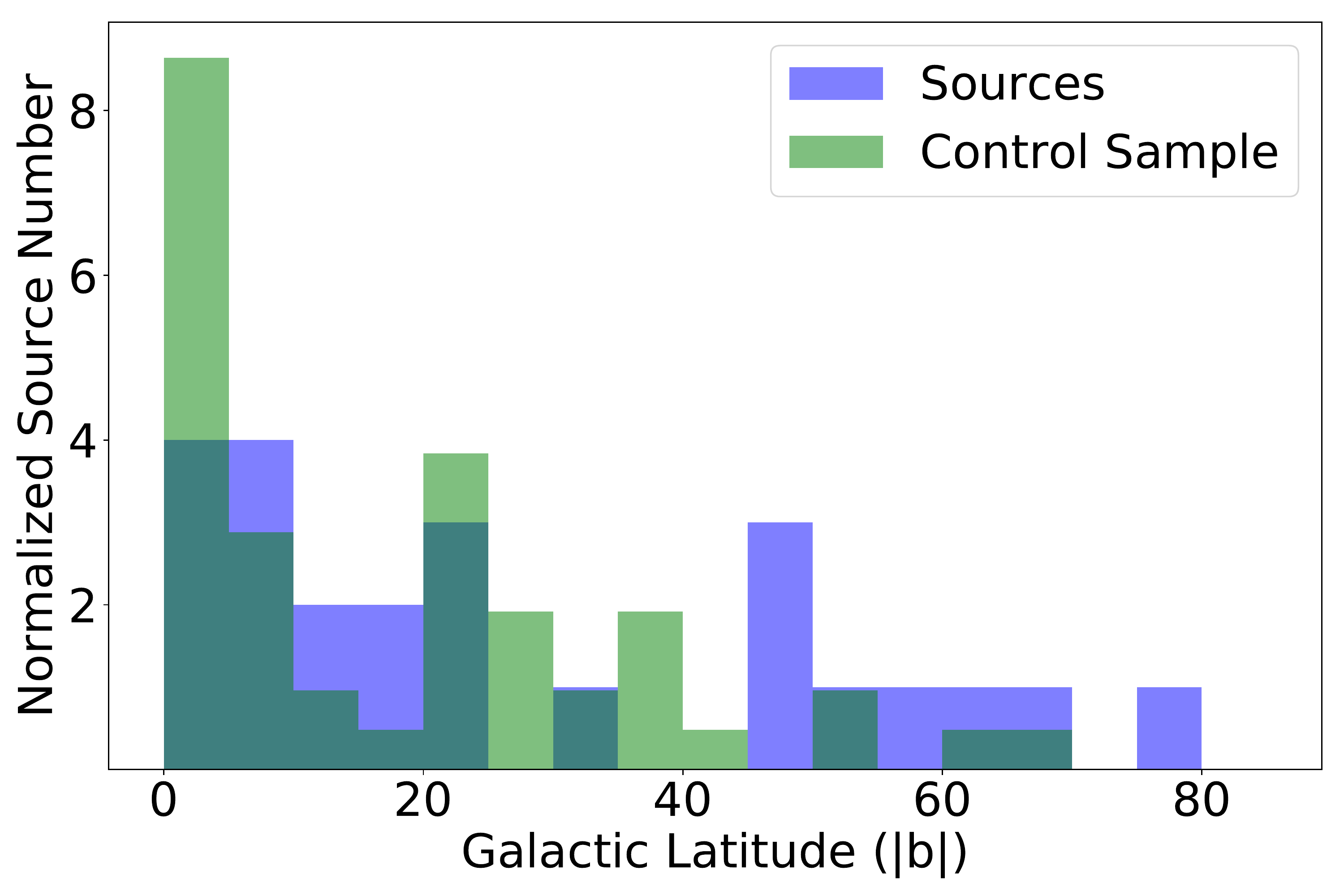}
\caption{The distribution of MSPs in galactic latitude for our source (blue) and control (green) samples. The number of control sources is divided by 50/24 so that the total number of sources in each population is equivalent.}
\label{latitude}
\end{figure}

\begin{figure}
\includegraphics[width=4.60in,angle=0]{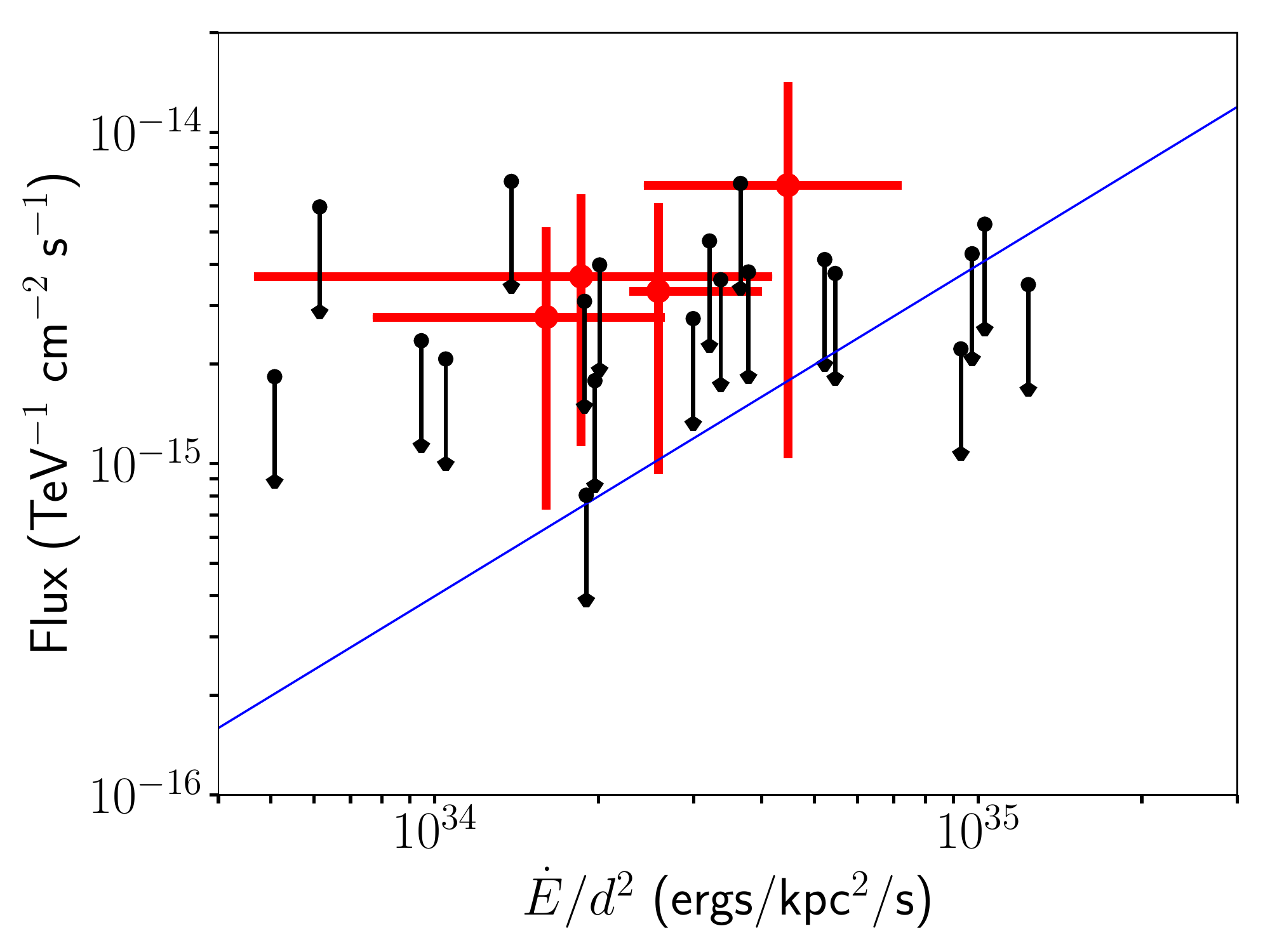}
\caption{The very-high energy gamma-ray fluxes yielded by the 2HWC Survey online tool in the directions of the 24 millisecond pulsars within the HAWC field-of-view with $\dot{E}/d^2 > 5 \times 10^{33}$ erg/kpc$^2$/s and that are not located near another 2HWC source. In utilizing this tool, we have adopted a Geminga-like degree of extension (see text for details). Sources detected at 2$\sigma$ are shown in red, while less-significant sources are shown with 2$\sigma$ upper-limits in black. The solid blue line denotes the expectation for pulsars with a Geminga-like efficiency for generating TeV halos.}
\label{main}
\end{figure}

It is worth checking whether the MSPs in our analysis are biased towards regions where galactic diffuse mis-modeling is more extreme, an effect which could potentially increase the number of observed high-TS sources. In Fig.~\ref{latitude} we plot the latitude distribution of both our MSP population and our control sample, finding that the bright MSP population is not biased towards the bright Galactic plane. In fact, the control sample includes more sources at low-galactic latitude, where diffuse mis-modeling is expected to be most severe.  This is not unexpected, as dimmer MSPs tend to be farther from the solar position, and thus are more likely to be at low galactic latitude. Of the four MSPs detected with (TS)$^{1/2} > 2.07$, J1939+2134 is observed in the galactic plane at ($b$~=~-0.29), while J1959+2048 ($b$~=~-4.70), J0613-0200 ($b$~=~-9.30) and J0030+0451 ($b$~=~-57.61) are found reasonably far from the plane.

In Fig.~\ref{main}, we plot the gamma-ray flux (at 7 TeV) obtained using the HAWC online tool (for the case of Geminga-like extension) verses the spindown flux of each MSP in our sample. We compare this to the flux predicted if each MSP transferred a Geminga-like fraction of its spindown power into its TeV halo (solid black line). We consider this dataset to be entirely compatible with the hypothesis that these MSPs are surrounded by TeV halos with characteristics that are broadly similar to those observed around Geminga and Monogem.

Alternatively, we note that two of the highest TS MSPs identified in this study (J1023+0038, J1959+2048) are either transient low-mass X-ray binary systems, or black widow pulsars. This could potentially suggest that such objects produce more TeV emission than the remainder of the MSP population~\cite{Venter:2015gga,Kisaka:2011sp}. On the other hand, J0030+0451 and J0613-0200 each appear to be isolated MSPs, while J0613-0200 is in a non-interacting binary system.

\section{Implications}
\label{implications}

The production of bright TeV halos by MSPs would have numerous consequences for high-energy astrophysics. As discussed in Sec.~\ref{model}, two separate physical effects must be present in order for MSPs to produce detectable TeV emission with a spatial extension similar to that of Geminga and Monogem. First, MSPs must convert a sizable fraction of their spindown power into very high-energy electron-positron pairs that escape from the pulsar magnetosphere. Second, the diffusion coefficient surrounding MSPs must be much smaller than that of the Milky Way's interstellar medium.

We begin by pointing out that the production of TeV halos by MSPs would significantly increase the number of TeV sources that are potentially detectable by upcoming HAWC observations. Although there are no MSPs with spindown luminosities large enough to be detected as 5$\sigma$ sources with current HAWC data (see Sec.~\ref{analysis}), it was estimated in Ref.~\cite{Linden:2017vvb} that 10~years of HAWC data could potentially detect TeV halo activity from any pulsar with a spindown flux exceeding $\sim$2\% of the Geminga value. We found 11 ATNF pulsars within the HAWC field-of-view\footnote{In that paper we utilized a slightly smaller declination cut of $-10^{\circ}$ to $50^\circ$.} with ages between 100-400~kyr that exhibited such a flux. This temporal cut was due to the necessity of separating TeV halo emission from possible hadronic contamination by the associated supernova remnant (on the low-end), and the rapid spindown evolution of young pulsars (on the high-end). Relaxing these cuts, and utilizing the larger field-of-view applied throughout this analysis, we find 45 pulsars in the ATNF catalog which exhibit fluxes at least 2\% as large as Geminga. In comparison, we find 20 MSPs with such a spindown power (see Table~\ref{table1}). This indicates that MSP-driven TeV halos could contribute substantially (or perhaps even dominate) the total number of observable TeV halos. 

As in the case of young pulsars, TeV halo searches may also potentially uncover a significant population of previously unknown MSPs, with radio beams that are not oriented towards Earth~\cite{Linden:2017vvb}. In the case of young pulsars, this population is likely to be substantial, as the typical pulsar with a spindown flux in the HAWC sensitivity range has a beaming fraction of approximately 30\%, indicating that there are approximately two ``invisible" pulsars capable of producing TeV halo emission for every radio pulsar~\cite{1998MNRAS.298..625T}. In the case of MSPs, the radio beaming fraction is likely to be significantly larger, decreasing the relative size of this invisible population~\cite{Ray:2012ms, 2008LRR....11....8L}. However, observational constraints on the MSP beaming angle currently depend on a comparison of the radio beaming fraction to the gamma-ray beaming fraction observed by instruments such as Fermi-LAT. TeV halo observations thus provide an independent handle on the size of the MSP beam. Additionally, the detection of TeV halo emission from an MSP without an observable radio flux would provide a new avenue for detecting nearby and cold pulsars. Such a system could potentially provide an extremely sensitive probe of the dark matter nuclear-scattering cross-section~\cite{Baryakhtar:2017dbj}. 

Finally, in addition to the potential detectability of individual TeV halos associated with MSPs, such sources may produce a significant background of diffuse TeV emission. While the total spindown power of MSPs is insignificant compared to that of radio-detected young-pulsars~\cite{2005AJ....129.1993M}, MSPs are capable of propagating far from the Galactic plane, making them an important contributor to high-latitude Galactic diffuse emission~\cite{FaucherGiguere:2009df, Abdo:2009ax, SiegalGaskins:2010mp}. In the case of the GeV-scale gamma rays observed by Fermi, this emission is inconsequential compared to the much brighter diffuse background dominated by extragalactic sources such as Active Galactic Nuclei and star-forming galaxies~\cite{Ackermann:2014usa}. At multi-TeV energies, however, most extragalactic sources are masked by gamma-ray attenuation, implying that the TeV halos surrounding MSPs could dominate the very high-energy diffuse emission observed at high-latitudes. Novel methods may be necessary to separate this emission from the cosmic-ray background contamination intrinsic to both Atmospheric- and Water-Cherenkov telescopes.


\section{TeV Halos and the Galactic Center GeV Excess}
\label{IG}

A number of groups have reported the detection of a bright and statistically significant excess of GeV-scale gamma rays from the direction of the inner Milky Way~\cite{Goodenough:2009gk,Hooper:2010mq,Hooper:2011ti,Abazajian:2012pn,Gordon:2013vta,Hooper:2013rwa,Daylan:2014rsa,Calore:2014xka,TheFermi-LAT:2015kwa,Karwin:2016tsw,TheFermi-LAT:2017vmf}. As the spectrum, morphology and intensity of this signal are each broadly consistent with that predicted from annihilating dark matter particles, such interpretations have received a great deal of attention (see, for example, Refs.~\cite{Ipek:2014gua,Izaguirre:2014vva,Agrawal:2014una,Berlin:2014tja,Alves:2014yha,Boehm:2014hva,Huang:2014cla,Cerdeno:2014cda,Okada:2013bna,Freese:2015ysa,Fonseca:2015rwa,Bertone:2015tza,Cline:2015qha,Berlin:2015wwa,Caron:2015wda,Cerdeno:2015ega,Liu:2014cma,Hooper:2014fda,Arcadi:2014lta,Cahill-Rowley:2014ora,Ko:2014loa,McDermott:2014rqa,Abdullah:2014lla,Martin:2014sxa,Berlin:2014pya,Hooper:2012cw}). In addition, astrophysical explanations of this emission have also been extensively discussed. At this time, the leading astrophysical interpretation of this observation is that it is generated by a large population of unresolved and centrally located millisecond pulsars~\cite{Cholis:2014lta,Petrovic:2014xra,Brandt:2015ula,Lee:2015fea,Hooper:2015jlu,Bartels:2015aea,Hooper:2016rap,Haggard:2017lyq}. If MSPs are surrounded by TeV halos, as suggested by the analysis presented in the previous section, TeV and radio observations of the Inner Galaxy could each be used to constrain or study the population of MSPs that is present in this region.


We begin by estimating the total spindown power of an MSP population capable of generating the observed GeV excess. To this end, we make use of the gamma-ray efficiencies of MSPs reported in the Fermi Second Pulsar Catalog (2PC)~\cite{TheFermi-LAT:2013ssa}. In order to mitigate the impact of observational bias, we consider only those MSPs with $\dot{E}/d^2 > 10^{34}$ erg/s/kpc$^2$, finding that this collection of 25 MSPs exhibits an average gamma-ray efficiency of $\langle \eta_{\gamma} \rangle \equiv {L_{\gamma}/{\dot E}} \simeq 0.12$, where $L_{\gamma}$ is defined as the gamma-ray luminosity integrated above 0.1 GeV. This sample only includes those pulsars with gamma-ray emission beamed in our direction, however.
As the emission from TeV halos is necessarily isotropic, all pulsars will contribute to the very high-energy flux, regardless of the geometry of their beams~\cite{Linden:2017vvb}. Taking these factors into account, we can write:
\begin{eqnarray}
L_{\rm GCE} &=& \langle \eta_{\gamma} \rangle \, f_{\rm beam} \, {\dot E}_{\rm tot},
\end{eqnarray}
where $f_{\rm beam}$ is the fraction of MSPs with gamma-ray emission beamed in our direction and ${\dot E}_{\rm tot}$ is the total spindown power of the central pulsar population. Comparing this to the luminosity of the GeV excess within a $0.5^{\circ}$ radius around the Galactic Center, $L_{\rm GCE} \simeq 2 \times 10^{36}$ erg/s ($>$0.1 GeV)~\cite{Calore:2014xka}, this requires the following approximate total spindown power among pulsars in this region (corresponding roughly to the $\sim$75 parsecs around the Galactic Center):
\begin{equation}
{\dot E}_{\rm tot} \sim 3.3 \times 10^{37} \, {\rm erg}/{\rm s} \, \times \bigg(\frac{0.12}{\eta_{\gamma}}\bigg)\,\bigg(\frac{0.5}{f_{\rm beam}}\bigg).
\end{equation}

\begin{figure}
\includegraphics[width=4.60in,angle=0]{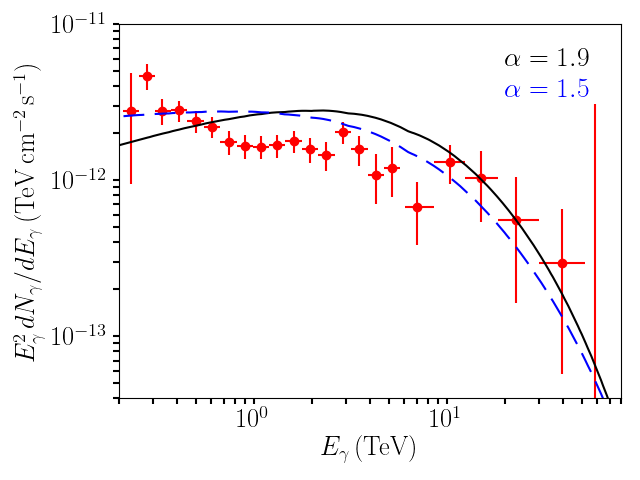}
\caption{The gamma-ray spectrum from the $0.2^{\circ}$ to $0.5^{\circ}$ (partial) annulus around the Galactic Center as reported by the HESS Collaboration~\cite{Abramowski:2016mir}. We compare this to the spectrum predicted from the TeV halos of a millisecond pulsar population that is capable of generating the Galactic Center GeV excess. Here we have adopted $B=10\,\mu$G, $\rho_{\rm IR}=6.0$ eV/cm$^3$, \mbox{$\rho_{\rm star}=6.0$ eV/cm$^3$,} $\rho_{\rm UV}=1.0$ eV/cm$^3$. The black (blue) curves correspond to $\alpha=1.5$, $E_{\rm cut}=35$ TeV ($\alpha=1.9$, $E_{\rm cut}=50$ TeV).}
\label{HESS}
\end{figure}

\begin{figure}
\includegraphics[width=4.60in,angle=0]{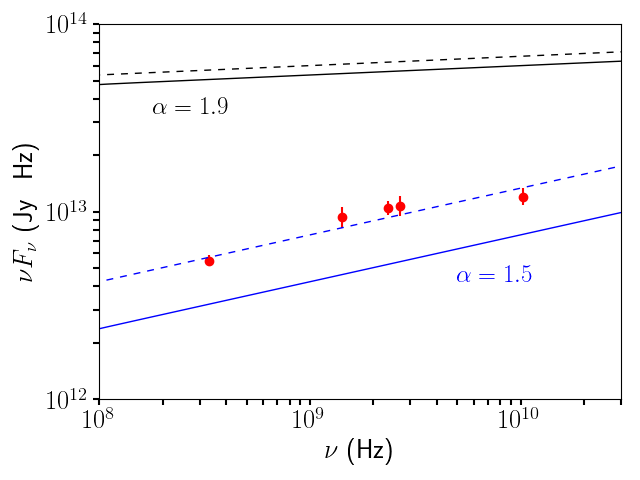}
\caption{The spectrum of synchrotron emission from the innermost $\pm 1^{\circ}\times \pm 3^{\circ}$ ellipse around the Galactic Center aligned along the Galactic Plane, assuming that sychrotron dominates over inverse Compton emission. The black (blue) curves correspond to $\alpha=1.5$, $E_{\rm cut}=35$ TeV ($\alpha=1.9$, $E_{\rm cut}=50$ TeV). Dashed (solid) lines are for the case of $B=10\, \mu$G ($B=100\, \mu$G). In each case, we have normalized the power in electrons/positrons to that required for MSPs to generate the Galactic Center GeV excess. The error bars represent the diffuse flux measured from the the same region~\cite{Crocker:2010xc}.}
\label{radio}
\end{figure}

If MSPs generate TeV halos with an efficiency similar to Geminga and Monogem (as suggested by the analysis of Sec.~\ref{analysis}), we can use the required total spindown power in the above equation to estimate the inverse Compton and synchrotron emission from this population of pulsars. In Fig.~\ref{HESS}, we plot the very high-energy gamma-ray spectrum from the TeV halos around an MSP population capable of generating the GeV excess, compared to the spectrum reported by the HESS Collaboration~\cite{Abramowski:2016mir}. To make this comparison, we additionally assume that the central MSP population has a spatial distribution that is consistent with the observed characteristics of the GeV excess, $n_{\rm MSP} \propto r^{-2.4}$, which we assume extends to a maximum of 3 kpc from the Galactic Center. We further describe the injected electron/positron spectrum by the parameterization $dN_e/dE_e \propto E_e^{-\alpha} \, \exp(E_e/E_{\rm cut})$ and adopt two parameter combinations which provide a good fit to the spectrum of Geminga ($\alpha=1.5$, $E_{\rm cut}=35$ TeV and $\alpha=1.9$, $E_{\rm cut}=50$ TeV)~\cite{Hooper:2017gtd}. We calculate the spectrum of the inverse Compton emission using the full Klein-Nishina treatment~\cite{Blumenthal:1970gc,Longair} and adopt a model for the radiation fields in the Inner Galaxy that consists of a superposition of following four components: $\rho_{\rm CMB}=0.260$ eV/cm$^3$, $\rho_{\rm IR}=6.0$ eV/cm$^3$, $\rho_{\rm star}=6.0$ eV/cm$^3$, $\rho_{\rm UV}=1.0$ eV/cm$^3$ and $T_{\rm CMB} =2.7$ K, $T_{\rm IR} =20$ K, $T_{\rm star} =5000$ K and $T_{\rm UV} =$20,000 K. In this figure, we also adopt a magnetic field strength of $B=10\, \mu$G.

From Fig.~\ref{HESS}, we see that if MSPs are responsible for generating the GeV excess, their TeV-scale emission should be expected to approximately saturate the emission observed by HESS from this region. The HESS Collaboration has argued, however, that this emission is correlated with the distribution of molecular gas, suggesting a hadronic origin for a non-negligible fraction of this flux~\cite{Abramowski:2016mir}, and significantly limiting the contribution produced by MSPs. Moreover, the high star-formation rate in the Galactic Center \cite{Figer:2008kf} indicates that young pulsars should also produce a significant population of TeV halos within the central molecular zone. In fact, a straightforward extrapolation of the TeV halo production efficiency and the star-formation rate indicates that TeV halos from young pulsars could dominate the TeV flux observed by HESS~\cite{Hooper:2017rzt}. One important caveat to this model, first pointed out in Ref.~\cite{Hooper:2017rzt}, concerns the possible disruption of electron acceleration in the PWN of young pulsars by the dense medium of the Galactic Center. However, the observation of TeV halos surrounding MSPs (even those near Earth), would argue against such an interpretation, because the electrons and positrons from MSPs are likely to be accelerated near the pulsar magnetosphere. Thus, it is difficult to imagine a scenario in which TeV halos from MSPs dominate the diffuse TeV gamma-ray flux near the Galactic Center, without there being a significant additional flux produced by young TeV halos.

One way to potentially reduce the TeV emission predicted in this scenario would be to imagine that most of the energy injected into very high-energy electron-positron pairs is emitted as synchrotron emission, rather than as inverse Compton. This, however, is constrained by radio observations of the Inner Galaxy. In Fig.~\ref{radio}, we plot the spectrum of synchrotron emission from the innermost $\pm 1^{\circ}\times \pm 3^{\circ}$ ellipse around the Galactic Center and aligned along the Galactic Plane, assuming an MSP population that generates the observed GeV excess and that a negligible fraction of the energy in very high-energy electrons/positrons goes into inverse Compton scattering. For each case shown, the predicted radio flux either dominates or exceeds the measured spectrum~\cite{Crocker:2010xc}. In calculating the synchrotron spectrum, we follow the approach described in Appendix B of Ref.~\cite{Strong:1998fr}, which is a pitch-angle averaged and relativistic approximation of the full electron synchrotron spectrum as initially calculated in Ref.~\cite{1970RvMP...42..237B}.

Figs.~\ref{HESS} and~\ref{radio} indicate that if MSPs do generate TeV halos with an efficiency similar to Geminga and Monogem, multi-wavelength observations of the inner Milky Way could be used to meaningfully constrain the size of the MSP population present in the inner Milky Way, and to test whether such sources could potentially generate the GeV excess observed by Fermi. We particularly look forward to future measurements of the very high-energy emission from this region by the Cherenkov Telescope Array (CTA).

\section{Summary and Conclusions}
\label{summary}

The HAWC Collaboration has recently reported the observation of bright and spatially extended multi-TeV emission from the young and nearby pulsars Geminga and Monogem. Furthermore, all indications are that such TeV halos are not restricted to these objects, but instead surround most or all young pulsars. In this study, we have utilized the HAWC data (through the 2HWC Survey online tool) in an effort to determine whether millisecond pulsars (MSPs) are also surrounded by TeV halos. To this end, we have studied 24 nearby and high spindown-power MSPs within HAWC's field-of-view, finding between 2.6 and 3.2$\sigma$ evidence that these sources are in fact surrounded by TeV halos. Furthermore, our analysis finds that the TeV efficiency of these sources (\ie the fraction of the total spindown power that goes into TeV emission) is approximately equal to that exhibited by Geminga and Monogem. 

If MSPs are confirmed to be surrounded by TeV halos at a high significance, this would have considerable implications for many facets of high-energy astrophysics. In addition to providing information about the acceleration and escape of electrons and positrons in MSP magnetospheres, this would lead us to expect that several dozen MSPs will ultimately be detected as high-significance sources by HAWC, including many of which do not have radio beams oriented in our direction. Furthermore, the TeV halos of unresolved MSPs could potentially dominate the diffuse emission at TeV energies, in particular at high galactic latitudes. 

Lastly, we point out that if MSPs are surrounded by TeV halos, the TeV and radio synchrotron emission from these sources could be used to constrain the population of MSPs present in the inner Milky Way. If MSPs do, in fact, generate the Galatic Center GeV excess, we find that this source population would also be expected to saturate or exceed the TeV and/or radio synchrotron emission observed from this region. We anticipate that future multi-wavelength observations will provide a powerful test of whether MSPs are responsible for the GeV excess.

\bigskip
\bigskip
\bigskip

\textbf{Acknowledgments.} We thank Petra Huentemeyer, Henrike Fleischhack and
Hao Zhou for helpful comments. DH is supported by the US Department of Energy under contract DE-FG02-13ER41958. Fermilab is operated by Fermi Research Alliance, LLC, under contract DE- AC02-07CH11359 with the US Department of Energy.

\bibliography{hawcmsp}
\bibliographystyle{JHEP}

\end{document}